\documentclass[aps,prl,twocolumn,groupedaddress,amsmath,longbibliography,superscriptaddress]{revtex4-2}

\usepackage{braket}
\usepackage{graphicx}
\usepackage{hhline}
\usepackage{ulem}
\usepackage{color}

\usepackage{xr}
\usepackage{amsmath}
\usepackage{amsthm}
\usepackage{comment}
\usepackage{amssymb}

\usepackage{mathtools}
\usepackage{multirow}

\newcommand{\Z}{\mathbb{Z}}
\newcommand{\parens}[1]{\left(#1\right)}

\newcommand{\Be}{\ensuremath{{^9}{\text{Be}}^{+} \,}}
\newcommand{\Mg}{\ensuremath{{^{25}}{\text{Mg}}^{+} \,}}
\newcommand{\Splus}{\ensuremath{S_{+} \,}}
\newcommand{\Sminus}{\ensuremath{S_{-} \,}}

\newcommand{\sz}{Q(1|0)}

\newcommand{\bsz}{(1-\sz)}

\newcommand{\en}{Y(0|1)}
\newcommand{\ez}{Y(1|0)}
\newcommand{\bez}{(1-\ez)}
\newcommand{\ben}{(1-\en)}

\newcommand{\az}{A(1|0,0)}

\newcommand{\baz}{(1-\az)}

\DeclareMathOperator*{\argmax}{arg\,max}
\DeclareMathOperator*{\argmin}{arg\,min}

\begin{document}

\title{High-fidelity indirect readout of trapped-ion hyperfine qubits}

\author{Stephen D. Erickson}
\email[]{stephen.erickson@colorado.edu}
\affiliation{National Institute of Standards and Technology, 325 Broadway, Boulder, CO 80305, USA}
\affiliation{Department of Physics, University of Colorado, Boulder, CO 80309, USA}

\author{Jenny J. Wu}
\affiliation{National Institute of Standards and Technology, 325 Broadway, Boulder, CO 80305, USA}
\affiliation{Department of Physics, University of Colorado, Boulder, CO 80309, USA}

\author{Pan-Yu Hou}
\affiliation{National Institute of Standards and Technology, 325 Broadway, Boulder, CO 80305, USA}
\affiliation{Department of Physics, University of Colorado, Boulder, CO 80309, USA}

\author{Daniel C. Cole}
\affiliation{National Institute of Standards and Technology, 325 Broadway, Boulder, CO 80305, USA}

\author{Shawn Geller}
\affiliation{National Institute of Standards and Technology, 325 Broadway, Boulder, CO 80305, USA}
\affiliation{Department of Physics, University of Colorado, Boulder, CO 80309, USA}

\author{Alex Kwiatkowski}
\affiliation{National Institute of Standards and Technology, 325 Broadway, Boulder, CO 80305, USA}
\affiliation{Department of Physics, University of Colorado, Boulder, CO 80309, USA}

\author{Scott Glancy}
\affiliation{National Institute of Standards and Technology, 325 Broadway, Boulder, CO 80305, USA}

\author{Emanuel Knill}
\affiliation{National Institute of Standards and Technology, 325 Broadway, Boulder, CO 80305, USA}
\affiliation{Center for Theory of Quantum Matter, University of Colorado, Boulder, CO 80309, USA}

\author{Daniel H. Slichter}
\affiliation{National Institute of Standards and Technology, 325 Broadway, Boulder, CO 80305, USA}

\author{Andrew C. Wilson}
\affiliation{National Institute of Standards and Technology, 325 Broadway, Boulder, CO 80305, USA}

\author{Dietrich Leibfried}
\email[]{dietrich.leibfried@nist.gov}
\affiliation{National Institute of Standards and Technology, 325 Broadway, Boulder, CO 80305, USA}


\date{\today}

\begin{abstract}
We propose and demonstrate a protocol for high-fidelity indirect readout of trapped ion hyperfine qubits, where the state of a \Be qubit ion is mapped to a \Mg readout ion using laser-driven Raman transitions.  By partitioning the \Be ground state hyperfine manifold into two subspaces representing the two qubit states and choosing appropriate laser parameters, the protocol can be made robust to spontaneous photon scattering errors on the Raman transitions, enabling repetition for increased readout fidelity.
	We demonstrate combined readout and back-action errors for the two subspaces of $1.2^{+1.1}_{-0.6} \times 10^{-4}$ and $0^{+1.9}_{-0} \times 10^{-5}$ with 68\% confidence while avoiding decoherence of spectator qubits due to stray resonant light that is inherent to direct fluorescence detection.
\end{abstract}


\maketitle

	Trapped ions are a leading platform for quantum information processing (QIP), exhibiting high fidelities in state preparation and measurement~\cite{myerson2008high,burrell2010scalable,harty2014high,christensen2020high,edmunds2021scalable,ransford2021weak,zhukas2021high}, single-qubit rotations~\cite{brown2011single, harty2014high}, and two-qubit entangling gates~\cite{ballance2016high, gaebler2016high, srinivas2021high, clark2021high}, as well as promising pathways to scalability~\cite{wineland1998experimental, kielpinski2002architecture, monroe2013scaling}.
	These high fidelity results have been demonstrated for systems of one or a few qubits at a time. 
	As QIP systems grow to tens of qubits or more~\cite{friis2018observation, wright2019benchmarking, arute2019quantum, bradley2019ten}, characterization of errors from control and readout must also include any undesirable crosstalk on neighboring ``spectator" qubits, which can be particularly harmful to fault-tolerant quantum error correction protocols~\cite{sarovar2020detecting, parrado2021crosstalk, hou2019experimental}.
	The type and magnitude of crosstalk errors varies across QIP platforms and architectures.
	
	In trapped ion QIP, one significant type of crosstalk is decoherence due to absorption of resonant photons by spectator ions; a single such photon absorbed by a nominally non-participating spectator ion will destroy any quantum information encoded in its internal state \cite{leibfried2004building, bruzewicz2019trapped}.
	This form of crosstalk has measurable impact on small circuit demonstrations that incorporate mid-circuit measurement~\cite{wan2019quantum, ryan2021realization, gaebler2021suppression}, though significantly lower crosstalk has been demonstrated in systems with well-designed, high-efficiency readout infrastructure \cite{crain2019high}.
    Methods of reducing resonant light crosstalk will be an essential requirement for large-scale fault-tolerant QIP with atomic qubits.
	Techniques such as quantum logic spectroscopy (QLS) \cite{schmidt2005spectroscopy}, where information about the state of a qubit is mapped via a shared motional mode to a different species of ion  for fluorescence detection, may be beneficial for this task because the photons used to read out the auxiliary ion species have negligible impact on any qubit ions \cite{barrett2003sympathetic, tan2016high}.
	QLS thereby avoids resonant light crosstalk, at the cost of mixed-species quantum logic.
	Ions of a second species are already used for sympathetic cooling in large quantum algorithms with trapped ions \cite{pino2021demonstration}.

    QLS-based readout has the potential to be quantum non-demolition (QND), where the state of the qubit is unchanged by the measurement process after the initial projection.  
    Perfectly QND measurements can be repeated arbitrarily many times to obtain high readout fidelity, even if the fidelity of a single repetition is low.
	In practice, measurements never fulfill this ideal, and the number of times they can be repeated while still improving the overall readout fidelity is limited. 
	Reference~\cite{hume2007high} demonstrated $6 \times 10^{-4}$ infidelity for reading out the state of an ${}^{27}\text{Al}^{+}$ optical clock qubit through repetitive QLS, ultimately limited by the 21 s lifetime of the ${}^{3}\text{P}_{0}$ qubit state.
	It was unclear whether this technique could be similarly useful for other ion species without the favorable electronic structure of ${}^{27}\text{Al}^+$ \cite{bruzewicz2019trapped}, for example hyperfine qubits that suffer from  off-resonant photon scattering errors during logic gates driven by Raman laser beams. 

\begin{figure}[tbp]
    \begin{center}
	\includegraphics[scale=0.95]{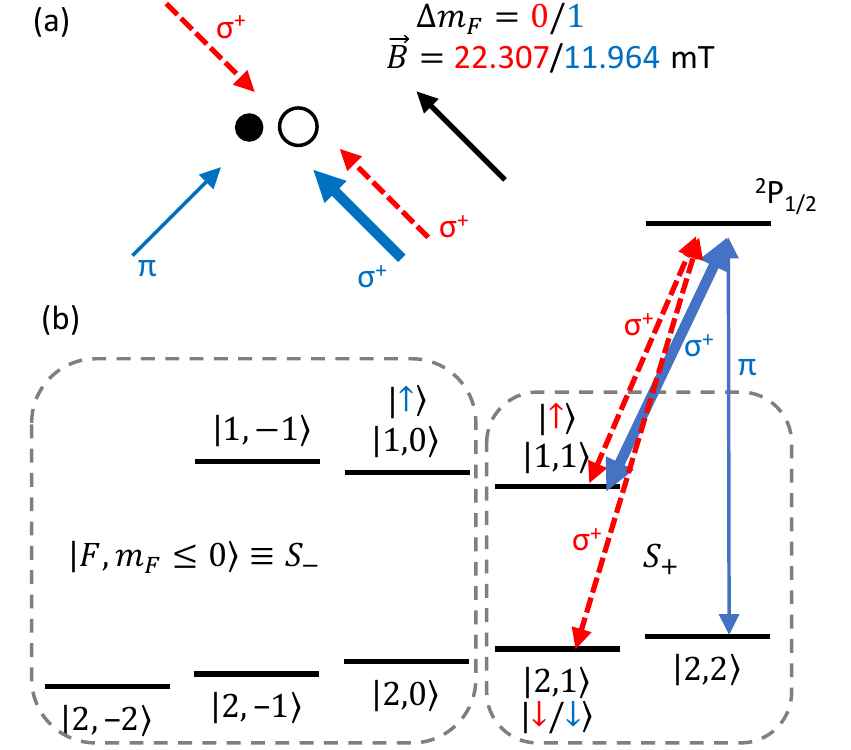}
    \end{center}
    \vspace*{-5mm}
    \caption[\Be Raman beam configuration and energy levels]{
		(a) Raman beam polarizations and geometries for containing spontaneously scattered qubit ion population using either of two possible configurations, color coded and labeled by the change in magnetic quantum number $m_F$ for the QLS transition ($\Delta m_F=0$ in dashed red, $\Delta m_F=1$ in solid blue). Magnetic fields are chosen to provide a field-insensitive qubit transition that can be driven with the same set of Raman laser beams. Small closed and large open circles represent the \Be qubit and \Mg readout ion, respectively. (b) Subspaces within the \Be $\,{^2\mathrm{S}_{1/2}}$ ground state manifold and associated field-insensitive qubit states (color coded) for either configuration. Line thickness varies between Raman beams to indicate relative intensities. 
	}
	    \vspace*{-5mm}
\label{fig:energy_levels}
\end{figure}
	
	In this Letter, we propose a technique to extend repetitive indirect readout to hyperfine qubits in a way that is resilient to off-resonant photon scattering errors, and we use it to demonstrate an order of magnitude reduction in indirect state readout infidelity relative to previous experiments with ions. 
	The key feature is to contain spontaneous photon scattering from the qubit Raman lasers within orthogonal subspaces by tailoring the laser beam intensities and polarizations, thereby ensuring that state-changing scattering events do not cause transitions between subspaces.
	Analogous subspace resilience to photon loss when reading out superconducting cavity qubits has been demonstrated \cite{elder2020high}.
	This technique is directly applicable to any trapped ion species with nuclear spin $\geq 3/2$, and can be adapted to other ion species that have extremely long lived excited states into which one qubit state can be transferred, e.g. the $^2 F_{7/2}$ state in Yb$^+$ ions \cite{ransford2021weak, edmunds2021scalable}.
	We propose two variants for reading out a ${}^{9}\text{Be}^{+}$ qubit using a co-trapped ${}^{25}\text{Mg}^{+}$ readout ion, labeled by the changes in magnetic quantum number $\Delta m_F$ in the ${}^{9}\text{Be}^{+}$ qubit that the Raman laser beams can drive (Fig. \ref{fig:energy_levels} (a)), and demonstrate the one that is compatible with our apparatus.
	The $^{2}\text{S}_{1/2}$ ground state manifold of $^9\text{Be}^+$, with energy eigenstates labeled $\ket{F, m_F}$, is divided into two orthogonal subspaces defined as $\Splus \equiv \{\ket{F,m_F \geq 1}\}$ and $\Sminus \equiv \{\ket{F,m_F \leq 0}\}$.
	The QLS scheme uses two-photon stimulated Raman transitions \cite{wineland1998experimental} that are designed to keep the qubit within a single subspace as shown in Fig. \ref{fig:energy_levels} (b), even in the presence of off-resonant Raman scattering errors.

	The $\Delta m_F = 0$ configuration, represented by dashed red in Fig. \ref{fig:energy_levels}, uses two $\sigma^+$-polarized Raman beams, ideally but not necessarily with equal intensity, to drive the $\ket{F=2,m_F=1} \leftrightarrow \ket{1,1}$ transition for QLS. 
	With this configuration, a good qubit choice is the same $\ket{2,1} \leftrightarrow \ket{1,1}$ transition that is first-order insensitive to magnetic field at an applied field of $|\vec{B}| \approx 22.307$ mT. 
	Before readout, $\ket{\uparrow}\equiv\ket{1,1}$ could be transferred to $\ket{2,-2}$ so that the population in $\ket{\uparrow}$ is moved to the $S_-$ subspace.
	With the use of composite pulse sequences and multiple shelving states in $S_-$, high shelving fidelity should be readily achievable, though imperfections in this process will add additional readout error.
    The choice of Raman beam polarizations closes the subspace $S_+$ under any off-resonant scattering processes due to the Raman beams, allowing for many QLS repetitions. 
	Transitions from \Splus to \Sminus require a Raman beam polarization error, and transitions from \Sminus to \Splus require multiple off-resonant scattering events given a successful initial transfer to $\ket{2,-2}$. 
	
	An alternative configuration, shown in solid blue in Fig. \ref{fig:energy_levels}(b), drives $\Delta m_F = 1$ transitions with a strong $\sigma^+$ and a weak $\pi$-polarized Raman beam.
	This is compatible with QLS on $\ket{2,2} \leftrightarrow \ket{1,1}$ and computation on the $\ket{2,1} \leftrightarrow \ket{1,0}$ qubit transition, which is first-order field-insensitive for $|\vec{B}| \approx 11.964$ mT and couples to the same Raman beam polarizations. 
	Consequently, prior to readout one would transfer $\ket{1,0} \rightarrow \ket{2,-2}$ and $\ket{2,1} \rightarrow \ket{2,2}$.
	The $\Delta m_F = 1$ variant retains most of the benefit of the $\Delta m_F = 0$ variant, except that the $\pi$-polarized Raman beam opens an additional pathway to transition from \Splus to \Sminus when scattering out of the $m_F=1$ states.
	Its intensity should be kept low to reduce this rate. 
	We consider the $\Delta m_F = 0$ variant superior to the $\Delta m_F = 1$ variant due to the former's improved subspace preservation and more efficient use of Raman beam power.
	However, due to experimental limitations on Raman beam geometry and magnetic field strength in our system, we demonstrate the method with the $\Delta m_F = 1$ variant. 

\begin{figure*}[htbp]
    \begin{center}
	\includegraphics[scale=0.92]{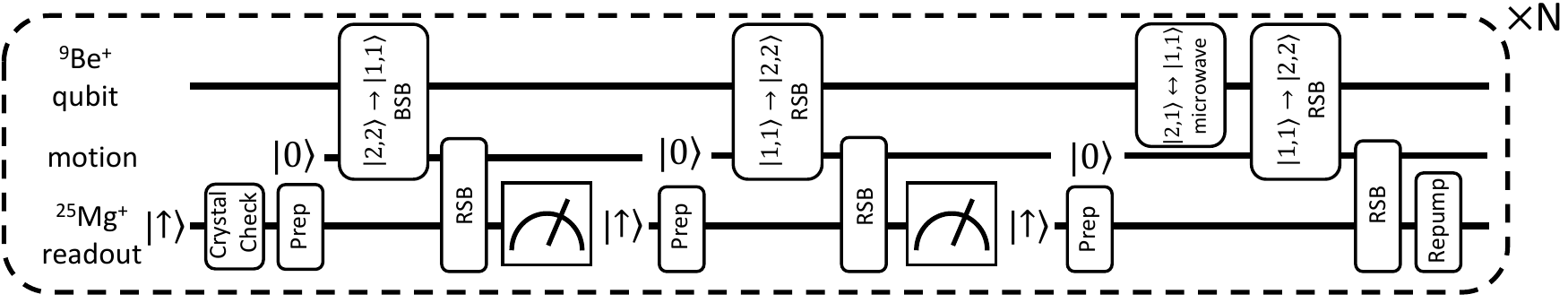}
    \end{center}
    \vspace*{-5mm}
    \caption[Quantum logic spectroscopy circuit]{
		Circuit for one repetition of the QLS protocol. Sidebands on the \Be qubit are used to inject phonons into the collective out-of-phase motional mode if in $S_+$. These phonons are then detected by the \Mg readout ion sidebands and subsequent fluorescence detections, yielding two bits of information per repetition. Leaked population from the \Be $\ket{2,1}$ state is then recovered. Each repetition has relatively low fidelity, but because the subspaces are preserved with high probability, high readout fidelity can be achieved by repeating the process $N$ times.
		}
    \vspace*{-5mm}
\label{fig:circuit}
\end{figure*}

	Since subspace preservation depends on the Raman scattering rate of the qubit ion, it is desirable to choose ion coupling parameters that minimize that rate, possibly even at the expense of single-repetition QLS fidelity. 
	This implies working with the highest feasible Raman beam detuning from excited states (e.g. red-detuning from the ${^{2}\mathrm{S}_{1/2}} \leftrightarrow {^{2}\mathrm{P}_{1/2}}$ transition in $^9\text{Be}^+$).
	Further benefit can be obtained by maximizing the \Be Lamb-Dicke (LD) parameter through choice of confining well, motional mode, and Raman beam wavevector difference $\Delta k$. 
	To this end, we operate on the ${}^{9}\text{Be}^{+} - {}^{25}\text{Mg}^{+}$ crystal axial out-of-phase (OOPH) mode at 2.91 MHz, with \Be and \Mg LD parameters of 0.37 and 0.097, respectively.
	Techniques based on the M{\o}lmer-S{\o}rensen interaction \cite{sorensen1999quantum,tan2015multi,bruzewicz2019dual,kienzler2020quantum, hughes2020benchmarking} could offer higher single-repetition QLS fidelity, but likely come at the cost of increased spontaneous Raman scattering from the qubit ion per QLS repetition.
	We therefore choose to use temperature-sensitive sideband-based QLS \cite{schmidt2005spectroscopy}, where information is mapped first from the qubit ion's internal state to the motion, and then from the motion to the readout ion.

	In our readout protocol, information is transferred from the \Be qubit ion, through the motional mode, to the \Mg readout ion with a qubit ion blue sideband (BSB, $\ket{2,2}\otimes\ket{n}\leftrightarrow\ket{1,1}\otimes\ket{n+1}$) or red sideband (RSB, $\ket{2,2}\otimes\ket{n}\leftrightarrow\ket{1,1}\otimes\ket{n-1}$) $\pi$-pulse followed by a readout ion RSB $\pi$-pulse.  
	After the transfer, the readout ion's state is determined using standard state-dependent fluorescence detection \cite{janik1985doppler}. 
	The scheme is designed to pump any population in $S_+$ into the state $\ket{2,2}$ and to leave any population in $S_-$ undisturbed.
	The full protocol is shown in Fig. \ref{fig:circuit} and detailed below.

    At the start of each experimental trial, we optically pump to the state $\ket{2,2}$, and if preparation in $S_-$ is desired, a sequence of microwave composite pulses is used to transfer the state to $\ket{2,-2}$.
    The pumping or transfer pulses can leave a small amount of erroneous population in the undesired subspace. 
    To further reduce that population, two sequences of the repetitive QLS protocol are performed back to back, the first of which heralds subspace preparation for the second.
    
	At the start of each QLS repetition, we perform a crystallization check by monitoring the fluorescence of the resonantly-excited readout ion to ensure that the ions are cooled to near the Doppler limit.  
	If this check fails, additional cooling is applied to the readout ion followed by a second crystalization check. 
    We then cool the collective motion through the readout ion and reprepare the readout ion.
    Next we apply a qubit ion BSB $\pi$-pulse that creates a phonon in the motional mode if the qubit ion is in $\ket{2,2}$ and transfers $\ket{2,2}$ to $\ket{1,1}$, having no effect on all other states.
	If the qubit ion was elsewhere in $S_+$ or anywhere in $S_-$, this operation ideally is off-resonant from any other allowable transition from the motional ground state, in which case no phonons are injected.
	A readout-ion RSB $\pi$-pulse and fluorescence detection then detects whether a phonon was injected.
	We again ground state cool via the readout ion and reprepare its internal state.
	Then we apply an RSB $\pi$-pulse to the qubit ion that creates a phonon if the qubit ion was in $\ket{1,1}$ and transfers $\ket{1,1}$ to $\ket{2,2}$.
	Again, the presence of a created phonon is detected using a readout ion RSB pulse and fluorescence detection.
	We then cool and reprepare the readout ion.

    The binary outcomes (``dark'' or ``bright'', or alternatively 0 and 1, respectively) of the two fluorescence detections depend on the qubit ion's initial state, taking nominal values of $(0,1)$ for initial state $\ket{1,1}$, $(1,1)$ for initial state $\ket{2,2}$, and $(0,0)$ for initial states in $S_-$ or $\ket{2,1}$.  
    Population in $\ket{2,1}$ can thus cause readout errors.
	
	To avoid remaining in $\ket{2,1}$, in the last stage of each repetition we use a microwave $\pi$-pulse to transfer any population in $\ket{2,1}$ to $\ket{1,1}$, and then to $\ket{2,2}$ with a RSB $\pi$-pulse.
    Given that scattering to $\ket{2,1}$ is expected to be a rare occurrence, rather than detecting whether a phonon was injected (which would indicate that the qubit had likely been in $\ket{2,1}$), we simply cool it away with an RSB pulse followed by repumping on the readout ion.  
    With this strategy, although population in $\ket{2,1}$ can cause an error during a single repetition, it is unlikely for any population in $\ket{2,1}$ to persist through multiple QLS repetitions.
	The $\Delta m_F=0$ variant would be done similarly, except with the roles of $\ket{2,2}$ and $\ket{2,1}$ reversed.
	
	This constitutes one full repetition of the QLS protocol, which can be repeated multiple times to increase the fidelity of the qubit readout.
	The number of useful repetitions is ultimately limited by the increasing cumulative probability of $S_+ \leftrightarrow S_-$ transitions due to spontaneous Raman scattering from the qubit ion.  
	We follow Ref. \cite{hume2007high} to determine the readout result after repeated rounds of QLS.
    Bayesian analysis is performed based on reference data to determine the posterior probability of being in a particular subspace given a sequence of QLS results, the most probable of which gives the result of the readout~\cite{supplementary}.
	The reference data consists of fluorescence detection binary outcomes obtained from single repetitions of QLS after preparing the qubit ion in $\ket{2,2}$ for \Splus or $\ket{2,-2}$ for $S_-$.

\begin{figure}[tbp]
    \begin{center}
	\includegraphics[keepaspectratio,width=86mm]{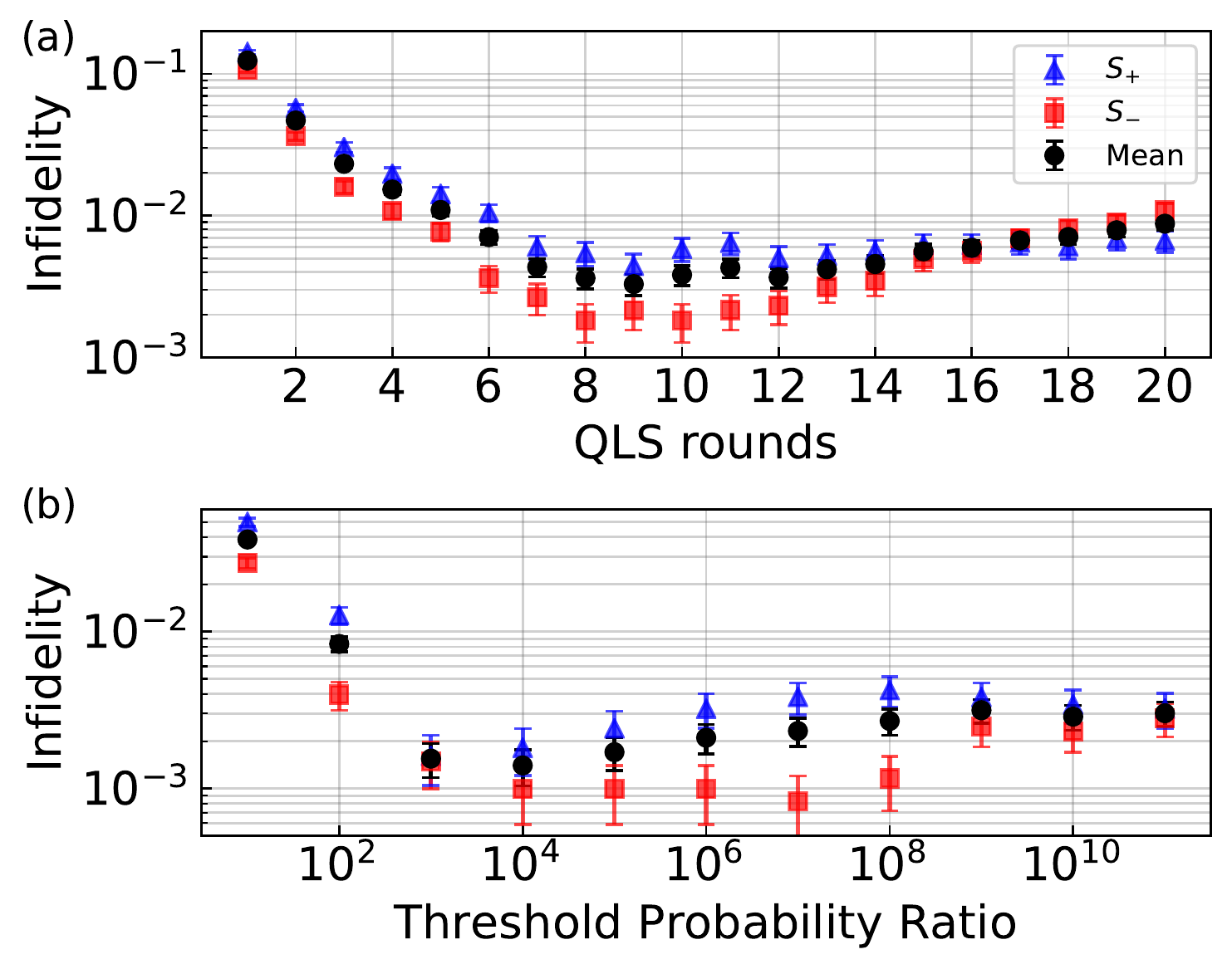}
    \end{center}
    \vspace*{-5mm}
    \caption[45 GHz test data]{
		Low-detuning test data showing (a) infidelity of the readout protocol vs. number of QLS repetitions per readout and (b) Infidelity vs threshold probability ratio for adaptive readout that repeats QLS until the threshold is reached. Blue triangles are used for \Splus infidelity, red squares for $S_-$, and black circles for their mean. Error bars are 68\% confidence intervals.
	}
    \vspace*{-5mm}
\label{fig:45GHz}
\end{figure}

    After heralding the qubit state as having been prepared in a given subspace with one sequence of the repetitive QLS protocol, a second sequence is applied without repreparing the qubit beforehand. 
    If this second readout disagrees with the first, then to lowest order either the second readout is in error or the first (heralding) readout changed the qubit subspace~\cite{supplementary}.
	We cannot distinguish between these two effects, so all readout infidelities we report are their sum (and hence an upper bound on each) to leading order.
	This leading order estimate is applied to the set of test data shown in Fig.~\ref{fig:45GHz}.
    We compute bounds on higher order corrections to the leading order estimates, and use them for our main results presented in Table~\ref{tab:detuning_data}~\cite{supplementary}.
    The corrections are small in comparison to the statistical uncertainties on the leading order estimates.
	
    Our demonstrations focus on analyzing the readout protocol itself, not how well our apparatus has been engineered to reliably implement it.
	For this reason we discard any experimental trials where the apparatus failed a status check, such as due to an optical cavity losing lock. 
	Furthermore, to remove the impacts of ion decrystallization and loss, we also discard any trials with failed crystallization checks on the readout ion throughout the QLS or for failed fluorescence checks on either species before/after each experimental trial.
	This method of selecting valid trials in real time could be used in near-term devices to increase readout fidelity at the expense of lowering algorithm execution rates.
	Prior to each trial we carry out a validation check by performing one repetition of QLS with the qubit prepared in each subspace in turn.
	We then track the fraction of the last 100 such validation checks that passed.
	If at any point either fraction falls below a preset threshold, the entire 100-trial window is discarded. 
	This guards against errors in the apparatus that are not caught by other validation checks, ensures that experiments where the apparatus fails are not erroneously counted as successfully reading out $S_-$, and protects against degradation of the QLS performance and, hence, the inferred fidelity of reading out $S_+$.
	
	To make readout infidelities easier to quantify for initial tests, we first apply the protocol to create a test dataset with Raman lasers 45 GHz red-detuned from the ${^2\mathrm{S}_{1/2}} \leftrightarrow {^2\mathrm{P}_{1/2}}$ transition and a 35 to 1 intensity ratio between the two beams.
	For comparison, 900 GHz detuning was previously used for high-fidelity entangling gates with \Be \cite{gaebler2016high}.
	The test data consist of 40 full repetitions of QLS per experiment, which we then analyze in post-processing. 
	The first $n$ repetitions are used to provide heralded state preparation and the next $n$ repetitions used to determine a readout fidelity, for various values of $n$ in the range $1\leq n \leq 20$. 
	The resulting infidelities for reading out either subspace, and their mean, are shown in Fig. \ref{fig:45GHz}(a). 
	Since each plotted point is derived from only the first $2n$ rounds of the same overall dataset, the plotted points and their error bars are partially correlated.
	The infidelity after only one QLS repetition is relatively high.
	However, it decreases steadily with additional repetitions, reaching a minimum mean infidelity of $3.3(6) \times 10^{-3}$ after nine repetitions.
	It then starts to gradually rise again due to the increasing cumulative probability of a spontaneous Raman scattering event in the qubit ion that changes the subspace.
	
	Not all fixed-length sequences of QLS repetitions reach the same posterior probabilities for being in either subspace.
	Likewise, the number of QLS repetitions required until the ratio of these probabilities exceeds some target value will vary depending on the sequence of QLS results. 
	We can therefore significantly reduce the average number of repetitions by actively tracking the posterior probability ratio of being in one subspace over the other, and stopping once the target ratio is reached~\cite{supplementary}.
	We refer to this as ``adaptive readout''~\cite{hume2007high,myerson2008high,crain2019high,todaro2021state}.
	Figure \ref{fig:45GHz}(b) shows the infidelity achieved for a range of threshold probability ratios using the same 45 GHz test dataset, analyzed adaptively in post-processing. 
	An infidelity of $1.4(4) \times 10^{-3}$ is reached for a $10^4$ probability ratio after an average of 3.47 repetitions, providing both an improvement in fidelity and a reduction in the average duration of the protocol compared to any fixed number of repetitions.

\begin{table*}[htbp]
\caption{
\label{tab:detuning_data}
	Observed infidelities of the adaptive readout protocol at 68\% confidence for a range of Raman beam detunings with their accompanying Raman beam intensity ratios, threshold posterior probability ratio, and mean number of QLS repetitions needed to exceed that ratio.
}
\begin{ruledtabular}
\begin{tabular}{lllllll}
Detuning (GHz) & Intensity Ratio	 & Threshold Ratio     & Mean Rounds & $S_{+}$ Infidelity    & $S_{-}$ Infidelity  \\ \hline
45             & 35:1	 & $10^4$	 & 3.55        & $2.7^{+0.6}_{-0.5}\times 10^{-3}$   & $2^{+3}_{-2}\times 10^{-4}$   \\
90             & 120:1	 & $10^7$	 & 5.13        & $8.1^{+4.2}_{-2.9}\times 10^{-4}$ & $2.6^{+2.8}_{-1.8}\times 10^{-4}$ \\
210            & 35:1	 & $10^7$	 & 5.92        & $2.5^{+1.3}_{-0.9}\times 10^{-4}$      & $0^{+1.8}_{-0} \times 10^{-5}$    \\
490            & 15:1	 & $10^9$	 & 8.55        & $1.2^{1.1}_{-0.6}\times 10^{-4}$    & $0^{+1.9}_{-0} \times 10^{-5}$   \\
\end{tabular}
\end{ruledtabular}
\vspace*{-5mm}
\end{table*}

	We also perform adaptive readout in real time on our experiment control field-programmable gate array (FPGA) for 45, 90, 210, and 490 GHz Raman detunings. The results are shown in Table \ref{tab:detuning_data}.
	To ensure that the infidelity only depends weakly on the threshold probability ratio, as observed in Fig. \ref{fig:45GHz}(b), we set the threshold conservatively high.
	This also guards against the possibility that drifts during the experiment relative to the single-round reference data could result in higher infidelity than the Bayesian analysis would otherwise suggest.
	At each detuning we made the Raman beam power imbalance as large as possible within the constraints of keeping sideband $\pi$-pulse durations $\tau$ within the range 5 $\mu$s $ \leq \tau \leq 40$ $\mu$s. 
	Shorter $\tau$ will drive carrier transitions off-resonantly, while longer $\tau$ makes the $\pi$-pulse fidelity more susceptible to drifts in the qubit or motional frequencies, for example due to fluctuating ac Stark shifts.  
	We also require the weak $\pi$-polarized beam to be strong enough to enable feedback stabilization of pulse envelopes.
	The real time data at 45 GHz align with those of the post-processed test dataset, and infidelity decreases with detuning, ultimately reaching $1.2^{+1.1}_{-0.6} \times 10^{-4}$ and $0^{+1.9}_{-0} \times 10^{-5}$ infidelity at $68\%$ confidence for $S_+$ and $S_-$, respectively, at 490 GHz detuning and a 15:1 intensity ratio ($1.2^{+2.39}_{-0.95} \times 10^{-4}$ and $0^{3.9}_{-0} \times 10^{-5}$ at $95\%$ confidence). 
	For comparison, the infidelity for reading out $S_+$ without the procedure to recover population from $\ket{2,1}$ is $4(2) \times 10^{-4}$, and the average single-repetition Raman scattering probability within $S_+$ is $5(1) \times 10^{-4}$, which was measured separately.
	
	At detunings of 210 and 490 GHz, the infidelity in reading out \Sminus is small and difficult to quantify; since multiple spontaneous Raman scattering events are required for population beginning in $\ket{2,-2}$ within \Sminus to scatter into $S_+$, the probability of leaving $S_-$ drops rapidly with the scattering rate.
	We observed no disagreements between the first and second readouts in roughly 100,000 experiments for reading out \Sminus in the 210 and 490 GHz datasets. 
	On the other hand, the probability of changing from $S_+$ to $S_-$ is given by a constant times the spontaneous Raman scattering rate.  
	This proportionality constant is much less than 1, and depends on the strong $\sigma^+$-beam polarization error and Raman beam intensity ratio.
	The probability to scatter out of \Splus could be reduced by using a qubit ion with larger nuclear spin because $S_+$ could include more states, and multiple scattering events would be required to exit the \Splus subspace. However, those additional states must be incorporated into the protocol by adding appropriate repumping steps (analogous to the repumping of $\ket{2,1}$).
	This difference between the \Splus and \Sminus scattering rates could be exploited to reach higher average readout fidelity by inverting the subspaces if initially in $S_+$. 
	Achieving that benefit requires that the error for exchanging subspaces is small compared to the infidelity in detecting either subspace.
	
	The duration of repeated QLS readouts, typically around 100 ms for the largest Raman detunings, sets a practical limit on the number of experimental trials, and thus the statistical power for quantifying the $S_-$ readout error. 
	However, the duration is dominated by ground state cooling, so QLS could be substantially sped up with alternative sub-Doppler cooling techniques, for example electromagnetically-induced transparency cooling \cite{roos2000experimental, lin2013sympathetic}.
	With the cooling duration minimized, ion fluorescence detection durations become significant, and can be reduced through Bayesian analysis that incorporates photon arrival times and can terminate early \cite{myerson2008high}.
	To achieve the highest fidelity, readout ion optical pumping and Doppler cooling durations were chosen very conservatively, but could likely be reduced in practice.

	In conclusion, we demonstrate indirect qubit subspace readout of trapped ions with an order of magnitude reduction in infidelity relative to previous work \cite{hume2007high}.
	The observed readout infidelities are competitive with the lowest readout infidelities (direct or indirect) of any qubit \cite{myerson2008high,burrell2010scalable,harty2014high,christensen2020high,elder2020high,edmunds2021scalable,ransford2021weak,zhukas2021high}. 
	The protocol extends repetitive quantum non-demolition measurements to hyperfine qubits in a way that is resilient to spontaneous Raman scattering.
	Alternatively, such scattering could be avoided by instead using near-field microwave gradients for spin-motion coupling \cite{wineland1998experimental,mintert2001ion,ospelkaus2011microwave,srinivas2019trapped}.
	The scheme also eliminates errors due to stray resonant laser light that can affect spectator qubits in large quantum processors. 
	We also suggest a $\Delta m_F = 0$ variant with balanced Raman beam intensities that will allow for more efficient use of available Raman beam power and fewer subspace-changing scattering events.  
	The technique can be used on any ion with nuclear spin $\geq 3/2$, and can be extended to ion species with nuclear spin $< 3/2$ by shelving to long-lived excited states, for example the $^2 F_{7/2}$ state in $^{171}$Yb$^+$ ions \cite{ransford2021weak, edmunds2021scalable}.
	An additional repump laser with well-controlled polarization may be necessary to clear metastable D states in some such ions.

\section{acknowledgments}

\begin{acknowledgments}
	We thank Yu Liu and Matthew Bohman for helpful comments on the manuscript. S.D.E, J.J.W., P.-Y.H., S.Geller, and A.K. acknowledge support from the Professional Research Experience Program (PREP) operated jointly by NIST and the University of Colorado. 
	S.D.E. acknowledges support from the National Science Foundation under grant DGE 1650115. 
	D.C.C. acknowledges support from a National Research Council postdoctoral fellowship.
	This work was supported by IARPA and the NIST Quantum Information Program.
\end{acknowledgments}

\bibliography{refs}

\clearpage
\pagebreak
\onecolumngrid

\begin{center}
	\textbf{\large Supplementary Information for ``High-fidelity indirect measurement of trapped-ion hyperfine qubits''}
\end{center}
\setcounter{equation}{0}
\setcounter{figure}{0}
\setcounter{table}{0}
\setcounter{page}{1}
\makeatletter
\renewcommand{\theequation}{S\arabic{equation}}
\renewcommand{\thefigure}{S\arabic{figure}}


This Supplementary Information describes how we estimate the qubit ion's subspace from QLS readout data and how we estimate the fidelity of QLS readout. Throughout the Supplementary Information, we use uppercase letters for random variables, and their lowercase counterparts for particular values of that random variable. We use the notation $\mathbb{P}(E)$ to indicate the probability of event $E$ occurring, and $\mathbb{P}(X=x)=\mathbb{P}(x)$ is the probability that random variable $X$ takes the value $x$.

\section{Subspace Estimation From Readout Data}

In this section we describe how reference data were collected to estimate the probabilities $\mathbb{P}\left(v|s\right)$ of measuring two-bit outcome $v$ when in subspace $s$ and how these estimates are used to determine readout outcomes with a Bayesian maximum a posteriori estimator (an estimator for the most probable subspace given the results of the readout).
To generate the reference data, 10,000 experimental trials were taken for each subspace at each Raman beam detuning. Each trial begins with preparing the qubit ion in one of the subspaces and then reading out that qubit.
Subspace preparation for reference data begins as described in the main text with optical pumping to $\ket{2,2}$ for $S_+$, followed by a series of microwave pulses to transfer population to $\ket{2,-2}$ for $S_-$.
To better approximate the distribution of populations during an arbitrary
repetition, instead of only the first, we apply one round of QLS as depicted in
Fig.~2 in the main text, except with the readout ion fluorescence measurements replaced with repumping. 
We then apply a second round of QLS exactly as depicted in Fig.~2, with each fluorescence measurement yielding a certain number of photons counted. The two photon-count values from the two measurements are converted to a two-bit outcome $v$ by assigning bit value 0 or 1 if the photon count is below or above a threshold.
From the 10,000 trials we obtain the estimates $\widehat{\mathbb{P}(v|s)}$ for each of the probabilities, where $\widehat{\mathbb{P}(v|s)}=f(v|s)/10,000$ and $f(v|s)$ is the number of times that outcome $v$ was observed when the qubit ion was prepared in subspace $s$. We use these estimates of the conditional probabilities obtained from known subspace preparations to analyze readout data obtained from uncertain subspace preparations.

To estimate unknown subspace preparations from a sequence of measurement results  obtained from multiple rounds of QLS, we use a Bayesian strategy with a uniform prior distribution $\mathbb{P}(S=S_-)=\mathbb{P}(S=S_+)=1/2$. 
Let $v_j$ be the measurement result obtained from the $j$\textsuperscript{th} round of QLS, and
$\vec{v}=(v_j)_{j=1}^M$ is a sequence of measurement results from $M$ QLS rounds.  After $\vec{v}$ has been observed, the a posteriori probability of being in subspace $s$ is
\begin{align}
\mathbb{P}\left(s|\vec{v}\right)=
\frac{\mathbb{P}\left(\vec{v}|s\right)\mathbb{P}(s)}{\sum_s\mathbb{P}\left(\vec{v}|s\right)\mathbb{P}(s)}.
\end{align}
Because the prior distribution is uniform, and we assume that each measurement is independent, 
\begin{align}
\mathbb{P}\left(s|\vec{v}\right)=
\frac{\prod_{j=1}^M\mathbb{P}(v_j|s)}{\sum_s\prod_{j=1}^M\mathbb{P}\left(v_j|s\right)}.
\label{eq:bayes}
\end{align}
Using the reference data, we have estimated each of the probabilities on the right hand side of Eq.~(\ref{eq:bayes}). Our estimate $\hat{s}$ of the state $s$ is the maximum a posteriori estimate, calculated from
\begin{align}
    \hat{s} = \argmax_s \frac{\prod_{j=1}^M\widehat{\mathbb{P}(v_j|s)}}{\sum_s\prod_{j=1}^M\widehat{\mathbb{P}\left(v_j|s\right)}}.
\end{align}

When using adaptive readout, we choose a target probability ratio $t$. After each round of QLS, we compute the ratio
$u=\prod_{j=1}^m\widehat{\mathbb{P}(v_j|S_-)}/\prod_{j=1}^m\widehat{\mathbb{P}(v_j|S_+)}$, where $m$ is the number of rounds observed so far and $(v_j)_{j=1}^m$ contains all outcomes from those $m$ rounds.  When $u>t$ or $u<1/t$, rounds of QLS are complete and the value $\hat{s}$ is reported.

\section{Readout Fidelity Analysis}

\subsection{Model description}
\begin{figure}[h]
  \centering
  \includegraphics[width=.5\columnwidth]{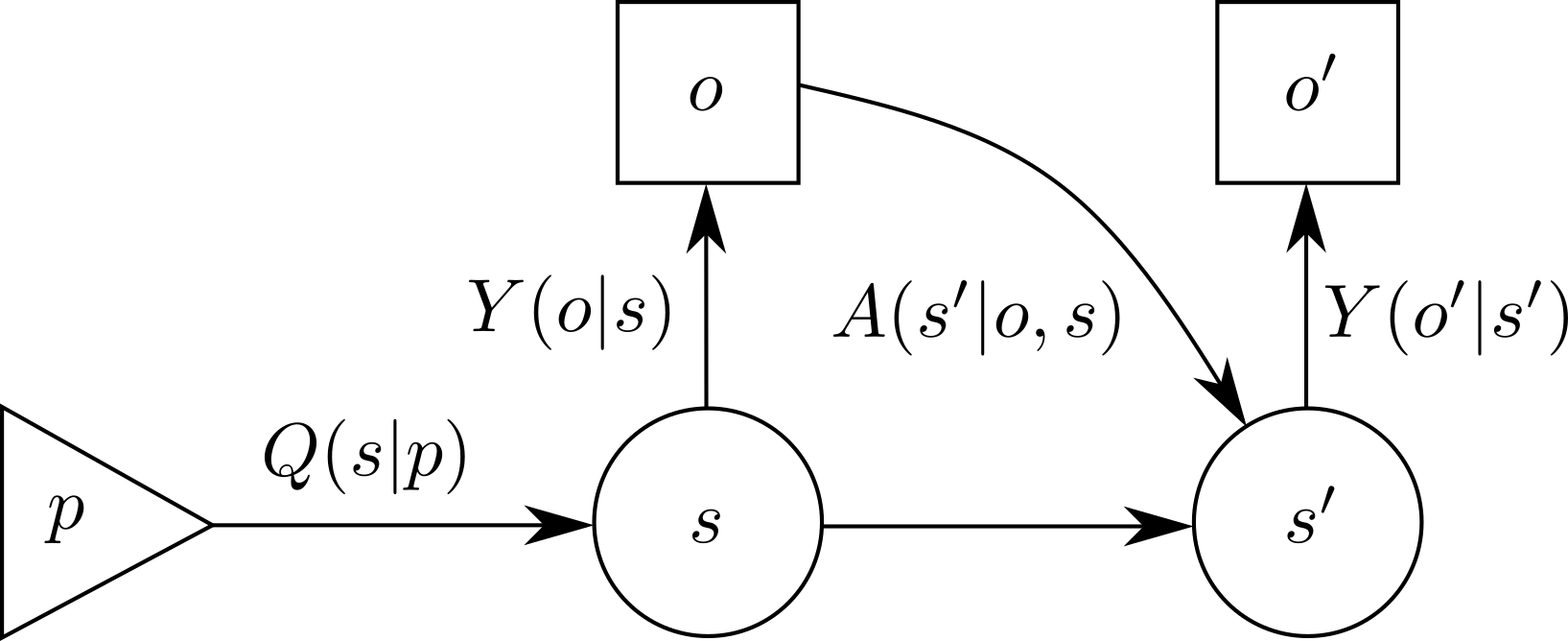}
  \caption{The coarse grained model used to estimate readout
  fidelity.}
  \label{fig:model}
\end{figure}
To estimate the readout 
infidelities and compute their confidence intervals, we 
perform two back-to-back readouts, each of which involves multiple repetitions of QLS. Because the readout process destroys any quantum coherence in the qubit, it suffices to use an
effective classical model for the subspaces and measurements. In this model, which is diagrammed in Fig.~\ref{fig:model}, there is a preparation in an intended state followed by two readout processes that occur sequentially in time and are assumed to be identical. This coarse-grained model abstracts the readout process, specifying only the probability of reading out an outcome given that the ion started the readout in a given subspace. At the end of the first readout, the ion is potentially in a different subspace, which is then the subspace that it starts in for the second readout.

In accordance with the classical
model, we define the random variables $S, S'$ taking values in $\Z_2 =\{0,1\}$ to indicate the subspace that the system is in when the first and second readouts begin, respectively. We use $\{0, 1\}$ instead of $\{S_+, S_-\}$ in this section for clarity in the calculations below. We define the random variables $O, O'$ taking values in $\Z_2$ for the first and second readout outcomes, respectively.
We use the symbol $P$ to denote the subspace that we intended to prepare before
the first readout. While it is not random, we use the same convention that the lowercase $p$ is a particular value for $P$.
Let $Q(s|p)$ denote the probability of having prepared subspace $s$ when intending to
prepare subspace $p \in \Z_2$. When the indices disagree, $Q$
describes the subspace preparation error. The readout process is described by
a conditional probability distribution $Y(o|s)$ of the probability
of reading outcome $o \in \Z_2$ when the system is in subspace $s
\in \Z_2$ at the beginning of the readout. Note that this description does not explicitly make reference to the actual state of the model during the readout process. At the end of the first readout, the system is in the same subspace that the system starts the second readout in, $s'$. We expect that $s'=s$, but in rare cases the system transitions to the other subspace, $s \neq s'$. Such events are captured by the conditional distribution $A(s'|o,s)$. We emphasize that our model makes no reference to the time at which a transition occurs.
We assume that the second readout is described identically to the first, by the distribution $Y(o'|s')$.
In sum, our model is specified by the readout probabilities $Y$, the state preparation error $Q$, and the transition probability $A$. There are eight independent parameters in total. 

Every sequence of events within the model is associated with a probability
\begin{align}
  \mathbb{P}\left( O=o, O'=o', S=s, S'=s'|P = p \right) &=
  Q(s|p)Y(o|s)A(s'|o,s)Y(o'|s').
  \label{eq:model}
\end{align}
However, only the probabilities $r(o',o|p) \coloneqq \mathbb{P}\left( O=o, O'=o'|P=p \right)$
can be estimated directly by the observed frequencies in the
experiment.
To express these quantities in terms of the model parameters we fix an outcome, and sum over
subspace sequences
\begin{align}
  r(o',o|p)  &= \sum_{s, s' \in
    \left\{ 0, 1 \right\}} Q(s|p)Y(o|s)A(s'|o,s)Y(o'|s').
  \label{eq:rdef}
\end{align}

The readout infidelities are represented by $Y(\neg x|x)$, where $\neg
0 = 1$ and $\neg 1 = 0$.
The system of equations Eq.~(\ref{eq:rdef})   cannot be solved directly for $Y(\neg x|x)$, so
instead we report estimates of $F(x) \coloneqq Y(\neg x|x) + A(\neg x|x, x)$. This choice is motivated by the fact that $r(\neg x,x|x) \approx
F(x)$ to lowest order in the various errors.  In order to make inferences about $F(x)$ without making the lowest-order
approximation, we compute upper and lower bounds for it in terms of the probabilities $r$. Because $A(\neg x|x, x) \geq 0$, an upper bound on $F(x)$ is also an upper bound on the readout error $Y(\neg x|x)$.

\subsection{Bounds on $Y(\neg x|x) + A(\neg x|x, x)$}
\label{sec:bounds}
In order to bound model parameters in terms of the observables $r$ we introduce the assumption that
\begin{align}
Y(x|x), Q(x|x), A(x|x, x) \ge c > 1/2\, ,
  \label{eq:cass}
\end{align}
for
some number $c$ to be chosen based on separate calibration data. We
then find subspace sequences that contribute to the expression for $r(o', o|p)$
given in Eq.~(\ref{eq:rdef}) that
contain only a single factor where the indices disagree, so that we can isolate
single model parameters in terms of $r$ and the parameter
$c$. For example, we have
\begin{align}
  r(1, 1|0) \ge Q(1|0)Y(1|1)A(1|1, 1)Y(1|1) \ge c^3 Q(1|0),
  \label{eq:firstbound}
\end{align}
dropping all other subspace sequences that contribute to the
sum. From this, we obtain a bound on the preparation error in terms of $r$ as
\begin{align}
  Q(1|0) \le r(1,1|0)/c^3.
  \label{eq:qbound}
\end{align}
Similarly, we obtain
\begin{align}
Q(\neg x|x) &\le \frac{r(\neg x, \neg x|x)}{c^3},
  \label{eq:standardboundq}\\
A(\neg x|x, x) + Y(\neg x|x) &\le \frac{r(\neg x, x|x)}{c^3}.
  \label{eq:standardbounday}
\end{align}

From separate calibration data, we expect that the quantities on the right hand sides in Eqs.~(\ref{eq:standardboundq}) and (\ref{eq:standardbounday}) are
small, so we take the model parameters on the left hand sides to be
small. We organize the expression $r(\neg x,
x|x)$ by order in these derived small quantities as
\begin{align}
  r(\neg x, x|x) = Y(\neg x|x) + A(\neg x|x, x) + h(Q,Y,A,x),
  \label{eq:firstorder}
\end{align}
where $h$ is the higher order contribution.
We want to calculate both lower and upper bounds for the quantity $F(x) = Y(\neg x|x) + A(\neg x|x, x)$ in
terms of just the observable quantities $r(w, y|z)$, and thus we want to find
upper and lower bounds for $h$. We accomplish this by using the bounds in
Eqs.~(\ref{eq:standardboundq}) and (\ref{eq:standardbounday}) as we now describe.

For concreteness we derive bounds on $F(0)$, which can be
extended
to bounds on $F(1)$ by the replacements $0 \leftrightarrow 1$. We expand the expression
in Eq.~(\ref{eq:rdef}) for $r(1,0|0)$ as
\begin{equation}
\label{eq:full_r_expr}
\begin{aligned}
r(1,0|0) =\;&\bsz\bez\Big(\az\ben + \baz\ez\Big) + \\
&\sz\en\parens{A(0|0,1)\ez + A(1|0,1)Y(1|1)}.
\end{aligned}
\end{equation}
To obtain an upper bound on $F(0)=Y(1|0)+A(1|0,0)$, we need only bound the
negative contributions in Eq.~(\ref{eq:full_r_expr}). The second term in Eq.~(\ref{eq:full_r_expr}) is positive and can be neglected. After expanding the
first term and again dropping positive terms, we have
\begin{equation}
r(1,0|0) \geq \Big(\ez + \az\Big) - b_-(Q,Y,A,x),
\end{equation}
with
\begin{equation}
\begin{aligned}
b_-(Q,Y,A,x) \equiv \Big(2&\az\ez + \sz(\ez+\az) + \az\en + \\ &\ez^2 + \sz\ez\az(\ez+\en)\Big).
\end{aligned}
\end{equation}
An upper bound for $b_-$ can be nicely grouped by adding an $\az^2$ term
\begin{equation}
  \begin{aligned}
b_-(Q,Y,A,x) <\;&\sz(\az+\ez) + \parens{\az+\ez}^2 + \\&\az\en + \sz\ez\az(\ez+\en).
  \end{aligned}
\end{equation}
Applying the bounds Eqs.~(\ref{eq:standardboundq},\ref{eq:standardbounday}) leads
to
\begin{equation}
b_-(Q,Y,A,0) < \frac{r(10|0)\left[r(10|0) + r(11|0) + r(01|1)\right]}{c^6}
+ \frac{r(11|0)r(10|0)^2\parens{r(10|0) + r(01|1)}}{c^{12}} \equiv u(r, c, 0),
\label{eq:upperbias}
\end{equation}
where we abbreviate $r(wy|z) = r(w, y|z)$.

We use an analogous technique to obtain a lower bound on $F(0)$. The second term in Eq.~(\ref{eq:full_r_expr}) is upper
bounded by $\sz\en$, so defining $b_+(Q,Y,A,0)$ to be an upper bound for all positive terms excluding $Y(1|0)
+ A(1|0,0)$ gives
\begin{equation}
  \begin{aligned}
b_+(Q,Y,A,0) \leq\;&\sz\en + \sz\parens{\ez^2 + \en\az + 2\ez\az}
+ \\&\ez\parens{\az\en + \ez\az}\, .
  \end{aligned}
\end{equation}

Using the same method of adding an $\az^2$ term to produce $(\ez+\az)^2$ and applying the bounds Eqs.~(\ref{eq:standardboundq},\ref{eq:standardbounday}) leads to
\begin{equation}
b_+(Q,Y,A,0) \leq \frac{r(11|0)}{c^3}\left[\frac{r(01|1)}{c^3} + \frac{r(10|0)^2}{c^6} + \frac{r(10|0)r(01|1)}{c^6} \right] + \frac{r(10|0)^2}{c^6}\parens{\frac{r(10|0)}{c^3}+\frac{r(01|1)}{c^3}}\equiv l(r, c, 0).
  \label{eq:lowerbias}
\end{equation}

As a result our final bounds are
\begin{equation}
r(\neg x, x|x) - l(r,c,x) \leq F(x) \leq r(\neg x, x|x) + u(r,c,x).
\label{eq:upperlowerschematic}
\end{equation}

\subsection{Reported quantities and confidence bounds}

\begin{table*}[htbp]
\caption{
\label{tab:full_detuning_data}
	Infidelities of adaptive readout
	for a range of Raman beam detunings. All quantities are computed with $c=.95$. The first subcolumn of each column are the values $f(\hat{r}, c=.95, x), g(\hat{r}, c=.95, x)$, with the upper value being $g$ and the lower being $f$. The 68\% subcolumn has the $\alpha/2 = (1-.68)/2$ significance upper bound for $g$ and the $\alpha/2$ significance lower bound for $f$, and similarly for the 95\% column. Note that the confidence intervals are asymmetric about the range $[f(\hat{r}), g(\hat{r})]$.
}
\begin{tabular}{|l|l|l|l|l|l|l|l|}
\hline
\multirow{2}{*}{Detuning (GHz)} 
     
& \multicolumn{3}{c|}{$S_{+}$ Infidelity}    & \multicolumn{3}{c|}{$S_{-}$
Infidelity}  \\ \cline{2-7}
& $f(\hat{r}), g(\hat{r})$  &68\% & 95\% & $f(\hat{r}), g(\hat{r})$ &68\% & 95\% \\ \hline
\multirow{2}{*}{45}             
	
&$2.75\times 10^{-3}$ &$3.40\times 10^{-3}$ &$4.02\times 10^{-3}$
& $2.04 \times 10^{-4}$ & $4.99\times 10^{-4}$  & $7.86\times10^{-4}$\\
&$2.72\times 10^{-3}$&$2.22\times 10^{-3}$&$1.79\times 10^{-3}$&$1.56\times 10^{-4}$&0&0\\\hline
\multirow{2}{*}{90}             
	 
& $8.13\times 10^{-4}$ & $1.23\times 10^{-3}$& $1.66\times
10^{-3}$& $2.66\times 10^{-4}$ & $5.44\times 10^{-4}$
& $8.19\times 10^{-4}$\\
&$8.08\times 10^{-4}$&$5.2\times 10^{-4}$&$3.3\times 10^{-4}$&$2.51\times 10^{-4}$&$8.1\times 10^{-5}$&$1.2 \times 10^{-5}$\\\hline
\multirow{2}{*}{210}             
	 
& $2.50\times 10^{-4}$ & $3.77\times 10^{-4}$& $5.07\times
10^{-4}$& $0$ & $1.83\times 10^{-5}$
& $3.74\times 10^{-5}$\\
&$2.49\times 10^{-4}$&$1.62\times 10^{-4}$&$1.02\times 10^{-4}$&$0$&$0$&$0$\\ \hline
\multirow{2}{*}{490}             
	 
& $1.16\times 10^{-4}$ & $2.33\times 10^{-4}$& $3.55\times
10^{-4}$& $0$ & $1.93\times 10^{-5}$
& $3.93\times 10^{-5}$\\
&$1.16\times 10^{-4}$&$5.2\times 10^{-5}$&$2.2\times 10^{-5}$&$0$&$0$&$0$\\ \hline 
\end{tabular}
\end{table*}

In this section, we describe the method used for constructing confidence
intervals on reported quantities. To describe the distinction between true
parameters and their estimates, we introduce the notation $\hat{a}$ to denote
an estimator of $a$. An estimator $\hat{a}$ is a function of the collected
data, while the parameter $a$ is not. 
In particular, we introduce the mean estimator for the probabilities $r(o',o|p)$,
\begin{align}
  \hat{r}(o', o|p)(n) &= \left.{n(o',o,p)}\middle/{\sum_{o',o}n(o',o,p)}\right.\!,
  \label{eq:meanestimator}
\end{align}
where the $n( o', o, p )$ are the number of observed experiments
with preparation $p$, first readout outcome
$o$ and second readout outcome $o'$, a total of eight numbers. We refer to the collection $\{n(o', o, p)| o', o, p \in \mathbb{Z}_2\}$ as the counts.
When the counts used to compute $\hat{r}$ are clear from context, we suppress
the $n$ dependence and write $\hat{r}(o', o|p)$.

In Table~\ref{tab:full_detuning_data}, we report $f(\hat{r}, c, x) \equiv \hat{r}(\neg x, x|x) - l(\hat{r},c,x)$, $g(\hat{r}, c, x) \equiv \hat{r}(\neg x, x|x) + u(\hat{r},c,x)$, 68\% and 95\% confidence bounds for $F(x)$, for $c=.95$. 
We note that while the distribution of the estimator $\hat{r}$ is not Gaussian, our choice of confidence levels to report is motivated by a comparison to this case. If the distributions were Gaussian, the 68\% and 95\% confidence intervals would be one and two standard deviations away from the mean, respectively.

We now describe how the confidence bounds in Table~\ref{tab:full_detuning_data} are computed.
We cannot directly estimate $F(x)$, but we can estimate the quantities $f(r, c,
x)$ and $g(r, c, x)$, which bound $F(x)$ according to Eq.~(\ref{eq:upperlowerschematic}). Note that $f(\hat{r}, c, x)$ is an unbiased
estimator of $f(r, c, x)$, and similarly for $g$. The uncertainties that we report are confidence lower bounds for $f(r, c, x)$ and confidence upper bounds for $g(r, c, x)$ at significance level $\alpha$, or equivalently at confidence level $1-\alpha$, for $0
< \alpha < 1$. These bounds are also confidence lower and upper bounds respectively for 
$F(x)$ at significance level $\alpha$. For $c\ge.9$, at the largest detuning
considered, we found for the counts obtained from the experiment
that $f(\hat{r}, c, x)$ and $g(\hat{r}, c, x)$ were within $10^{-5}$  of
each other and of $\hat{r}(\neg x, x|x)$ for both $x=0$ and $x=1$.
For all measured detunings, these differences are small compared to the 68 \% confidence intervals.

Next, we describe the procedure for computing
confidence upper bounds, and note that the procedure for computing confidence
lower bounds is similar. 
Our strategy is to derive a confidence upper bound for $f(r, c, x) \equiv
r(\neg x, x|x) + u(r, c, x)$, so that by Eq.~(\ref{eq:upperlowerschematic}), the
derived confidence upper bound for $f(r, c, x)$ is a confidence upper bound for $F(x)$. All confidence bounds that we report are conservative, in the sense that they are designed to contain the true value of estimated parameters with probability $\geq 1-\alpha$.

For a quantity $q$ that we wish to estimate, we use the notation
$\overline{q}(\alpha; n)$ to denote
a confidence upper bound for $q$ at significance level $\alpha$, for observed
counts $n$.
For each $w, y, z \in \mathbb{Z}_2$, we can estimate $\overline{r(w,
y|z)}(\delta; n)$
from the counts using a Clopper-Pearson~\cite{clopper1934use}
interval.
To compute $\overline{f(r, c, x)}(\alpha; n)$ for a desired $\alpha$, our strategy is then to separately obtain confidence upper bounds for $u$ and
$r(\neg x, x|x)$, then to use the union bound to combine them to obtain
a confidence upper bound for $f$ as
\begin{align}
  \overline{f(r, c, x)}(\alpha; n)_\gamma &= \overline{r(\neg x,
  x|x)}(\alpha-\gamma; n)
  + \overline{u(r, c, x)}(\gamma; n),
  \label{eq:funion}
\end{align}
for a parameter $0 < \gamma < \alpha$, to be chosen independently of the data, that determines the
confidence levels at which we estimate confidence upper bounds on $r(\neg x,x|x)$ and $u$.
The quantity $u$
itself depends on three different $r(w, y|z)$'s, and we use the union bound to
estimate a confidence upper bound on $u$ from the confidence bounds of the $r(w,
y|z)$'s. Therefore, if we compute a confidence upper bound for each of the relevant $r(w, y|z)$'s at
significance level $\beta$, we can compute a confidence upper bound for $u$ at significance level $3\beta$.
Since $u$ is monotonic increasing in $r$, we can write
\begin{align}
  \overline{u(r, c, x)}(3\beta; n) = u(\overline{r}(\beta; n), c, x).
  \label{eq:uunion}
\end{align}
 Therefore to get a total significance of
$\alpha$ for the confidence upper bound of $r(\neg x, x|x) + u(r, c, x)$, we
estimate a confidence upper bound of significance $\alpha-3\beta$ for $r(\neg x,
x|x)$. That is, we take $\gamma = 3\beta$ in Eq.~(\ref{eq:funion}).
Confidence lower bounds are computed similarly by using the lower bound $l$ in
Eq.~(\ref{eq:lowerbias}) instead of $u$.

\subsection{Choosing $\beta$}
  For a fixed $\alpha$, any choice of $\beta \in (0, \alpha/3)$ gives a legitimate
  confidence bound.
  However, we would like a choice that yields the tightest possible confidence intervals for
  counts close to what we expect. To determine such a choice for $\beta$, we use a set of artificial training counts,
  detailed in Eq.~(\ref{eq:trainingdata}), which is chosen to be close to what we expect based on separate
  calibration data.
We denote by $N_x = \sum_{o', o}n(o', o, x)$ the total
number of readouts taken with preparation $x$.
The set of artificial training counts $n_t$ that we used
has $N_0 = 25000, N_1 = 100000$, $\hat{r}(\neg x, x|x)(n_t) = \hat{r}(x, \neg x|x)(n_t)
  \approx 10^{-4}$, and $\hat{r}(x, x|\neg x)(n_t) \approx 10^{-2}$ for each $x \in
  \mathbb{Z}_2$. These choices correspond to the counts
  \begin{equation}
  \label{eq:trainingdata}
  \begin{aligned}
   n_t(1, 1, 1) &= 98980\\
   n_t(0, 1, 1) &= 10\\
   n_t(1, 0, 1) &= 10\\
   n_t(0, 0, 1) &= 1000\\
   n_t(1, 1, 0) &= 250\\
   n_t(1, 0, 0) &=   2\\
   n_t(0, 1, 0) &=   2\\
   n_t(0, 0, 0) &= 24746,
  \end{aligned}
  \end{equation}
  where any differences
  from the values $\hat{r}(n_t)$ stated in the previous paragraph are due to rounding, so that
  $N_0, N_1$ remained fixed.  The counts in Eq.~(\ref{eq:trainingdata}) are completely determined by the choices of $N_0$, $N_1$, and the values of $\hat{r}(n_t)$ and the rounding; no random sampling is performed.
  For counts $n$, we define the $\beta$ that gives the tightest confidence upper
  bound to be
  \begin{align}
    \beta^*(n, \alpha, c, x) \equiv \argmin_{\beta\in (0, \alpha/3)} \overline{f(r, c,
  x)}(\alpha; n)_{3\beta}.
    \label{eq:betaopt}
  \end{align}
  We computed for $c = .95$ with $\alpha=0.68$ that
  $\beta^*(n_t, .317/2, .95) = .001$.

  \subsection{Sensitivity to the choice of $\beta$}
  In order to ensure that our choice of $\beta$ does not adversely affect the
  size of the
  confidence intervals when the real counts are not equal to the artificial training counts, we perform a check of additional artificial counts $\left\{
  n^{(i)} \right\}$. These counts are chosen such that the associated $\hat{r}(n^{(i)})$ lie on a Cartesian grid with ten
uniformly spaced values in each direction, up to rounding. Specifically, 
$\hat{r}(1, 1|0)(n^{(i)})$ lies in the range $\left[ 3\times
10^{-3}, 3\times 10^{-2} \right]$, $\hat{r}(1,0|0)(n^{(i)})$ lies in the range $[3\times10^{-5}, 3\times10^{-4}]$, and $\hat{r}(0, 1|1)(n^{(i)})$ lies in the range $[3\times 10^{-6}, 3\times10^{-5}]$.
The total number of experiments for each preparation for all $n^{(i)}$ are fixed at $N_0 = 25000$ and $N_1=100000$.
For ease of computation for this check, we do not range over the other directions, and instead take counts such that $\hat{r}(0,1|0)(n^{(i)}) = \hat{r}(1,0|0)(n^{(i)})$, $\hat{r}(0,0|1)(n^{(i)}) = \hat{r}(1, 1|0)(n^{(i)})$,
and $\hat{r}(1, 0|1)(n^{(i)}) = \hat{r}(0, 1|1)(n^{(i)})$.
For each set of artificial counts $n^{(i)}$ taken from this grid we compare the
confidence bounds $\overline{f(r, c, x)}(\alpha; n^{(i)})_{3\beta_0}$ obtained by
using $\beta_0=.001$ to confidence bounds $\overline{f(r, c, x)}(\alpha; n^{(i)})_{3\beta^*_i}$ obtained by using
a $\beta^*_i= \beta^*(n^{(i)}, \alpha, c, x)$ that is optimized for that particular $n^{(i)}$. This comparison is based on
the distance from the interval edge to the point estimate and is normalized by
the size of the point estimate to give a percent loss. Specifically, the percent loss is
\begin{align}
  d(\beta_0, n, \alpha, c, x) \equiv \frac{\overline{f(r, c, x)}(\alpha; n)_{3\beta_0}
  - \overline{f(r, c, x)}(\alpha;
n)_{3\beta^*(n, \alpha, c, x)}}{\hat{r}(\neg x, x|x)}.
  \label{eq:lossfuncdef}
\end{align}

For $c = .95$ and for $\alpha = .317/2, \alpha = .045/2$, we find that the maximum percent loss incurred by
our choice of $\beta_0$ over these intervals is less than $10\%$. Seeing that the
loss is relatively small over the range of concern, we deemed $\beta_0 = .001$
to be an acceptable choice for computing the confidence intervals.

\end{document}